\documentclass{article}
\usepackage{epsfig}
\usepackage{amscd,amsmath}
\usepackage{amssymb}
\begin{document}
\setcounter{equation}{0}
\renewcommand{\theequation}{\thesection.\arabic{equation}}
\vspace{2cm}
\title{Field Dynamics on the Light Cone : Compact versus Continuum
Quantization}
\author{S. SALMONS$^{(1)}$, P. GRANG\'E$^{(1)}$ E. WERNER$^{(2)}$\\
(1)Laboratoire de Physique Math\'ematique et The\'orique\\
Universit\'e Montpellier II \\
F-34095, Montpellier, Cedex 05 - FRANCE \\
(2)Instit\"ut f\"ur Theoretische Physik\\
D-93040, Universit\"at Regensburg - GERMANY}
\date{} 
\maketitle

\vspace{2.5cm}

\begin{abstract}
Compact canonical quantization on the light cone (DLCQ) is examined
in the limit of infinite periodicity lenth L. Pauli Jordan
commutators are found to approach continuum expressions with marginal
non causal terms of order $L^{-3/4}$ traced back to the handling of
IR divergence through the elimination of zero modes. In contrast
direct quantization in the continuum (CLCQ) in terms of field
operators valued distributions is shown to provide the standard
causal result while at the same time ensuring consistent IR and UV
renormalization.
\end{abstract}
\vfill
\hrule\vskip10pt
\noindent{\small PM 99/05, Feb. 1999\hfill
}
\vfill\eject
\setcounter{equation}{0}

\section{Introduction}

Light front quantization has emerged as an important tool in the study of
non-perturbative aspects of field theories [1]. However a major problem in this approach
resides in the infrared behaviour of the continuum theory. Recently this issue was
clarified on the basis of a mathematically well defined procedure [2]. In the early
attempts to deal with these infrared problems, discretized light front quantization
(DLCQ) [3] has played an important role. The popularity of DLCQ resides in the easy and
conceptually simple treatment of the infrared regularisation : zero modes in the
expansion of the fields were simply eliminated and later on understood as the
LC-counterpart of the non-trivial ground state of equal-time (ET) quantization. The
study of critical phenomena in the framework of effective theories requires using a
continuum version of the quantum field theory on the light front. Indeed critical
points, critical exponents etc... are accessible only from a complete knowledge of the
cut-off dependence of the critical mass, which can only be given by the continuum
theory. In DLCQ the limit of infinite periodicity length L cannot be achieved in a
straightforward manner without further insights both on the handling of zero modes and
restoration of covariance and causality in the limiting process [4]. Our approach [2] was
to propose a genuine continuum treatment (CLCQ) in which fields are treated as operator
valued distributions, thereby leading to a well defined handling of ultravioled and light
cone induced infrared divergences and of their renormalization. We focussed in [2] on the
comparison of the critical coupling in the LC and ET-framework, showing that the continuum
non-perturbative LC-approach is no more complex than usual perturbation theory in lowest
order. The LC-critical coupling is in essential agreement with the RG-improved
perturbative result at fourth order. Here we want to report on a detailed comparison
between DLCQ and CLCQ treatments of important quantities like Pauli-Jordan commutator
functions, which, due to necessary concision and lack of space, could not be treated
therein.

 In Section 2 we recall the DLCQ and CLCQ Fock expansion of the field operators and the
resulting Pauli-Jordan field commutators. A detailed comparison of their behaviour in
terms of the periodicity length $L$ (e.g. intrinsic  cut-off $\Lambda$) is made in section 3
where the issue of covariance and causality is also discussed in the limiting process $L
\to \infty$. Some conclusions and perpectives are presented in section 4.

\setcounter{equation}{0}

\section{DLCQ and CLCQ field operators and commutators}

DLCQ was introduced [3] to resolve the zero mode problem. This mode is clearly isolated
from other modes and its explicit treadtment results in a "zero-mode constraint", the
solution of which carries the non-perturbative aspects of the theory. In the particle
sector, periodic boundary conditions are imposed, $L$ being the periodicity length,
leading to the usual Fock expansion. Restricting to $1+1$ dimension\footnote{For
massive field the
IR-problematics can be discussed independently of higher dimensionalities.} the particle
sector field writes 

\begin{equation} \phi(x) = \sum^{\infty}_{n=1} {1 \over \sqrt{4\pi n}} [a_n e^{-ik_nx} + a^+_n
e^{ik_nx}] \end{equation}
with
$$[a_n, a^+_m] = \delta_{n,m} \ \ , \ \ n, m \geq 1 ;$$
and
$$k_n = {n \pi \over L} \ \ , \ \ n \in \mathbb Z.$$
The CLCQ approach relies on the introduction of field operator-valued distributions
defined with respect to $C^{\infty}$-test functions with compact support [5].
%%%%%%%%%%%%%%%%%%%%%%%%%%%%%%%%
Apart from formal considerations there exists a fondamental physical argument which
demonstrates that it is compelling to treat the field amplitudes in the distributional
sense in order to guarantee that the LC quantization procedure by itself is correct.
Due to the hyperbolic form of the LC-Laplacian, initial field values have to the
prescribed on characteristics, i.e. on $x^+ = 0$ and $x^- = 0$. In order to be able to
transform this characteristic value problem into a problem with periodic boundary
conditions, test functions $f(p^+,p^-)$ have to be introduced with the property [6]

\begin{equation}\lim_{p^+ \to 0} {1 \over p^+} f(p^+, {m^2 \over p^+}) = 0\end{equation}
(see eq. (3.20) of ref. [6]).

This is exactly what happens automatically with the test functions defined below.
Condition (2.2) ensures, as discussed in detail in ref. [6], that the field values on
the characteristic $x^- = 0$ become dependent quantities and, as a consequence, the
quantization can be performed prescribing boundary values for $x^+ = 0$ at $x^- = - L$
and $x^- =  L$, where $L \to \infty$.
%%%%%%%%%%%%%%%%%%%%%%%%%%%%%%%%%%%%%%%%%%%%%%%%%%%%%
 The field
can be expressed in a chart independent way as a surface integral over a manifold,
thereby showing that the ultraviolet (UV) behaviour on the Minkowski manifold
dictates the UV and IR behaviour on the LC manifold.
This is due to the regularisation
properties of the test function which are automatically transfered from the first to the
second case.

In this context the field writes

\begin{equation}\phi_{LC}(x) = \int^{\infty}_0 {dp^+ \over 4 \pi p^+} [a(p^+) e^{-ip.x} +
a^+(p^+) e^{ip.x}] f_{LC} (p^+, \hat{p}^-(p^+)) \end{equation}

with
$$ [a(p^+), a^+(p\prime ^+)] = 4 \pi p^+ \delta(p^+ - p\prime^+).$$
In (2.2) $\hat{p}^-(p^+)$ stands for the on-shell condition $m^2/p^+$ and $f_{LC}$ is
the test function in momentum space which falls off with all its derivatives sufficiently
fast as a function of the Minkowski arguments  $p_0, p_z \ \ (p^+ = {1 \over 2} (p^0 +
p^3) \ \ , \ \ p^- = {1 \over 2} (p^0 - p^3))$. Its behaviour as a function of $p^+$ is
discussed in [2] : the singular behaviour of ${1 \over p^+}$ in (2.2) is completely
damped out by the behabiour of $f_{LC}$ for $p^+ \to 0$, eliminating $p^+=0$ as an
accumulation point. The ensuing renormalization is independent of the particular choice
of $f_{LC}.$

We examine first the Pauli Jordan communator $\Delta(x) = [
\phi(x) , \phi(0)]$ evaluated at $x^+ = 0$. In the DLCQ case on finds

%%%%
\begin{eqnarray} 
\Delta_{DLCQ} (x^+ = 0, x^-) & = &  \sum ^{\infty}_{n = - \infty \ , \ \neq 0} {1 \over 4
 \pi n} e^{-i {n \pi x^- \over L}} \nonumber \\
& = & - {i \over 4} [sign (x^-) - {x^- \over L}], \end{eqnarray}

where $sign (x) = \pm 1 \ \ \ if  \ \ x \gtrless 0, sign (0) = 0.$

Within CLCQ, with $\hat{f}(p^+) \equiv f_{LC} (p^+, \hat{p}^-(p^+))$, the corresponding
expression writes 

\begin{equation} 
\Delta_{DLCQ} (x^+ = 0, x^-) = - {i \over 2 \pi} \int_0^{\infty} {dp^+ \over p^+}
 \hat{f}^2
 (p^+)\sin(p^+ x^-). \end{equation}
The test function $\hat{f}$ is strictly one in the interval $[{1 \over \Lambda} , 
\Lambda -
{1 \over \Lambda}]$, varies between $0$ and $1$ in the intervals $[0, {1 \over \Lambda}]$
and $[\Lambda - {1 \over \Lambda}, \Lambda]$, and is zero outside.

\vfill\eject

\setcounter{equation}{0}

\section{Comparison of the DLCQ and CLCQ Pauli Jordan commutators}

The behaviour of $g_A(x^-) = 4i \Delta_{DLCQ} (x^+ = 0, x^-)$ is sketched in Figure 1.
\begin{center}
\epsfig{file=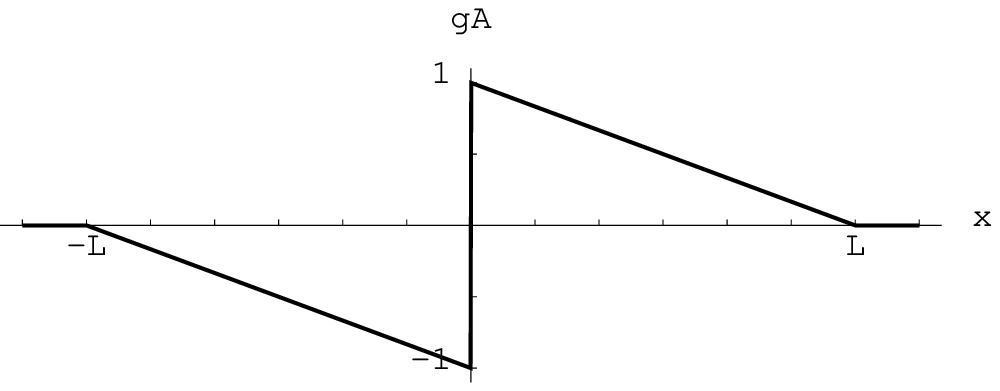}\\
 Fig.1: the DLCQ function $g_A(x^-)$.
\end{center}
To evaluate $g_B(x^-) = 4i \Delta_{CLCQ} (x^+ = 0, x^-)$, we choose
%%%%%%%%%%%%%%%%%%%%%%%%%%%%%%%%%
\begin{equation}
f(p) = \left \{ 
\begin{array}{ll}
1 - \exp \displaystyle[{1 \over \Lambda^2 p^2-1} + 1] \ \ \ & 0 \leq p  < {1 
\over \Lambda} \\
1 \ \ \ \ \  & {1 \over \Lambda} \leq p \leq \Lambda - {1 \over \Lambda} \\
1 - \exp \displaystyle[{1 \over \Lambda^2 (p - \Lambda )^2 - 1} + 1]  & 
\Lambda - {1 \over
\Lambda} < p \leq \Lambda \\
0 \ \ \ \ \ & p > {\Lambda}
\end{array} \right.
\end{equation} 
  
%%%%%%%%%%%%%%%%%%%%%%%%%%%%%%
with $\Lambda = 100$, and calculate $g_B(x^-)$ numerically. The results are  plotted
in Figure 2 at three different spatial scales.
\begin{center}
\hspace*{5.cm}\epsfig{file=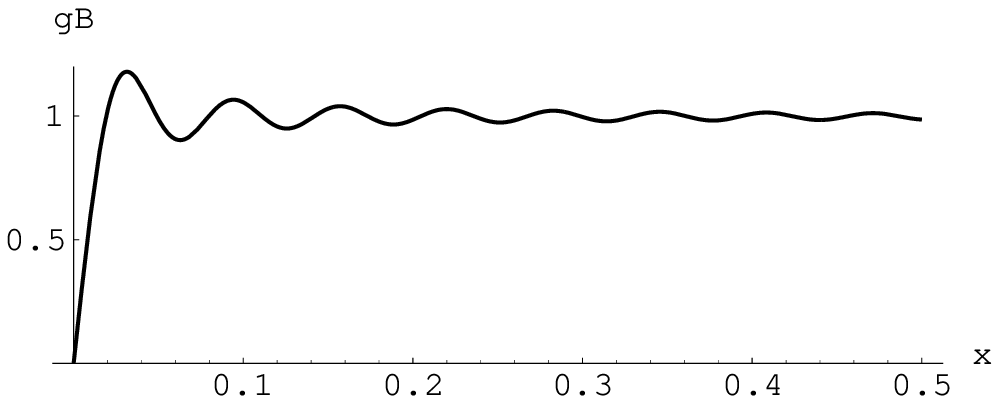,height=2.5cm} \vspace{-1.3cm}\\
 \hspace*{2.cm}
\epsfig{file=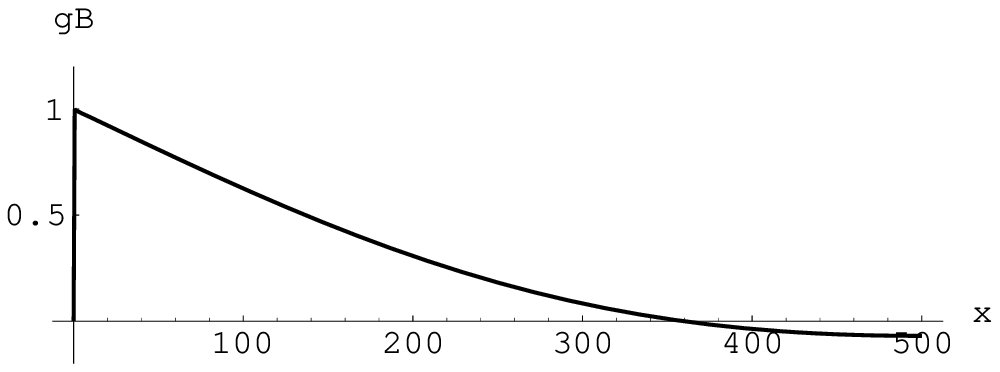,height=2.5cm} \vspace{-1.5cm}\\
\hspace*{-1.cm}\epsfig{file=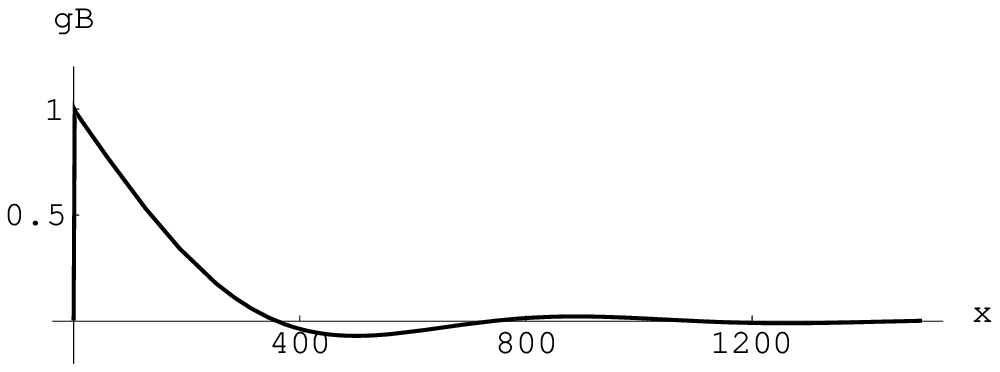,height=2.5cm}\\
 Fig.2: the CLCQ function $g_B(x^-)$ at different spatial scales.
\end{center}
 Near the origin $g_B(x^-)$ rises to 1 over distances shorter
 with increasing $\Lambda$. It is followed by an oscillatory fall-off with an average
 slope in $1 \over \Lambda$, corresponding to the straight line of $g_A(x^-)$ in DLCQ.
 Finally for large values of
$x^- \ \ (\geq 10 \Lambda) \ \ g_B(x^-)$ remains oscillating around zero.

Hence in both cases the decay zone and the asymptotic region where $g(x^-)$ is null or
quasi-null, reflect the elimination of the zero mode, $n = 0$ for DLCQ and a halo around
$p^+ = 0$ for CLCQ. However it is the presence of the UV-regularisation in CLCQ which is
responsible for the smeared out rise near $x^- = 0$ and small short wave length 
oscillations
for small $x^-$, at variance with DLCQ where no such regularisation is present. 
Clearly the n-summation can be arbitrarily cutt-off to deal with the UV-divergence 
but the approach to the continuum is not under control since
the limiting procedure of infinite cutt-off and infinite periodicity length compatible with causality is not known. To discuss these points we examine now the commutator
for space or time like separation.

For DLCQ we have
\begin{equation}
\Delta_{DLCQ}(x^+, x^-) = - {i \over 2 \pi} \sum^{\infty}_{n=1} {1 \over n}
 \sin [{n \pi x^- \over L} + {1 \over 4} {m^2 Lx^+ \over \pi n}] \end{equation}
and for CLCQ the corresponding expression is

\begin{equation}
\Delta_{CLCQ}(x^+, x^-) = - {i \over 2 \pi} \int_0^{\infty} {dp^+ \over p^+}
 \sin [{1 \over 4} {m^2 x^+ \over p^+} + p^+x^-] \hat{f}^2(p^+). \end{equation}

The integral in (3.3) is convergent even if $\hat{f} = 1$ everywhere and a straightforward change
of the integration variable shows that $\Delta_{CLCQ}$ depends only
on the product $x^+x^-$. The limit $\Lambda \to \infty$ can be taken safely with 
the result

\begin{equation}
\Delta_{CLCQ}(x^+, x^-) = - {i \over 4} [sign
(x^+) + sign (x^-)] J_0(m \sqrt{x^+x^-}), \end{equation}
which is the correct causal covariant expression, with $J_0(x)$ the Bessel function of
order zero.

Clearly for $\Delta_{DLCQ}$ the limit $L \to \infty$ cannot be taken before the sum is
carried out, as the sinus becomes ill-defined. As shown in Appendix A this limit
requires some care. Using eq.(A.19) one finds

$$\left. \Delta_{DLCQ}(x^+, x^-) \right|_{L \to \infty} = - {1 \over 4} [sign (x^+) 
+ sign (x^-)] J_0(m\sqrt{x^+x^-})$$
\begin{equation}
 + {i \over 2 m L x^+} \sqrt{x^+x^- + 2Lx^+ sign (x^+)} J_1 (m \sqrt{x^+x^-
+ 2Lx^+ sign(x^+}) + O(L^{-5/4}) \end{equation}

Hence the causal covariant expression is retrieved in the limit $L \to \infty$. However
 the marginal non causal term in $J_1$ in eq.(3.4) originates from the elimination of 
the zero mode in the infinite sum of eq.(3.2) (for $x^+ = 0$ it is just $ {i \over 4}
 {x^- \over L}$, cf. eq(2.4)). Its disappearance as $L \to \infty$ indicates that in 
the continuum the infrared problems would remain at variance with CLCQ. Thus in DLCQ,
 $L$ has to be kept finite to achieve IR regularisation, at the expense of the appearance
of a causality violating term of order $(L^{-3/4})$. Due to the regularisation properties
of the test functions, the situation in CLCQ is far more satisfactory since 
the approach provides a well defined handling of UV and IR-divergences and of their 
renormalization.

%%%%%%%%%%%%%%%%%%

\section{Conclusion}

It has been shown that dynamical properties of LC-quantized scalar fields whose basic
manifestation is in the Pauli-Jordan commutator function differ essentially if
quantized in DLCQ or in the continuum. DLCQ on a finite interval yields causality
violating terms being proportional to $L^{-3/4}$ which come in addition to the frame
independent result of the continuum theory. Unfortunately this does not mean that the
two versions coincide in the limit $L \to \infty$ since in this limit the infrared
regularisation of DLCQ is lost.

To conclude we want to add a remark concerning the LC-lattice method introduced by
Destri and de Vega [7] and elaborated by Faddeev and coworkers [8]. This approach works
on a LC-space-time lattice. The basic building blocks of field dynamics being causal
transfer matrices between neighbouring points along light-like directions, problems
with causality are avoided by construction in this discretization scheme. However the
main argument in favour of this approach lies in the integrability properties in
closest connection to those of the continuum. 

\vspace{1cm}

{\bf ACKNOWLEDGEMENTS}\\

We thank Drs. G. Mennessier for clarifying discussions and N. Scheu for pointing out reference[4]. This work has been completed under Nato Grant $n^o$ CRG920472

\section{References}
\begin{description}
\item[1.] "New Non-Perturbative Methods and Quantization on the light cone" Les Houches
Series, Vol. 8 (1998), Editors : P. Grang\'e, H.C. Pauli, A. Neveu, S. Pinsky,
 E. Werner. EDP-Sciences, Springer.
\item[2.] P. Grang\'e. P. Ullrich, E. Werner, Phys. Rev. D57, 4981, (1998).
\item[3.] H.C. Pauli, S.J. Brodsky, Phys. Rev. D32, 1993, (1985)\\
 S.J. Brodsky, H.C. Pauli and S.S. Pinsky Phys. Report 301, 299, (1998).
\item[4.] N. Scheu, Ph.D. thesis Universit\'e Laval (Quebec), hep-th/9804190.
\item[5.] L. Schwartz, Th\'eories des distributions, Hermann, Paris 1966\\
A.M. Gelfand and G.E. Shilov, Generalized functions, Academic Press, New-York 1967.
%%%%%%%%%%%%%%%%%%%%%%%
\item[6.] T. Heinzl, E. Werner, Z. Phys. C62, 521, (1994).
\item[7.] C. Destri, H.J. de Vega, Nucl. Phys. B290, 363, (1987), H.J. de Vega,
preprint LPTHE-PAR 94/26 hep-th/9406135.
\item[8.] L.D. Faddeev, A.Yu. Volkov, hep-th/9710039.
%%%%%%%%%%%%%%%%%%%%%%%%%%%%%
\end{description}
\setcounter{equation}{0}
\renewcommand{\theequation}{A.\arabic{equation}}
\section*{Appendix A}

\vspace{1cm}

In this appendix we derive the expression of the periodic
Pauli-Jordan function in the limit of infinite priodicity length L.
 
Consider the periodic distribution with period $ \lambda = {2\pi \over K}$ 
\begin{equation} f(x) = \sum^{\infty}_{n=-\infty} C_n e^{inKx} \end{equation}
and the class of $C^{\infty}$-test function $\varphi(x)$ with the properties :
\begin{equation} \{x \in [0,1] \ ; \   \varphi(x) + \varphi(x-1) = 1 \  ;  \ 
 \varphi(0) = 1,
\varphi(1) = 0,\left. {d^P \varphi(x) \over dx^P} \right|_{x=1} = 0 \ \ \  \forall p \geq 1 \}.
\end{equation}
This constitutes a decomposition of unity since by construction
\begin{equation} \sum^{\infty}_{p=-\infty} \varphi(x+p) = 1 \ \ , \ \ \forall x. 
\end{equation}
%
%\begin{figure}
\begin{center}
\epsfig{file=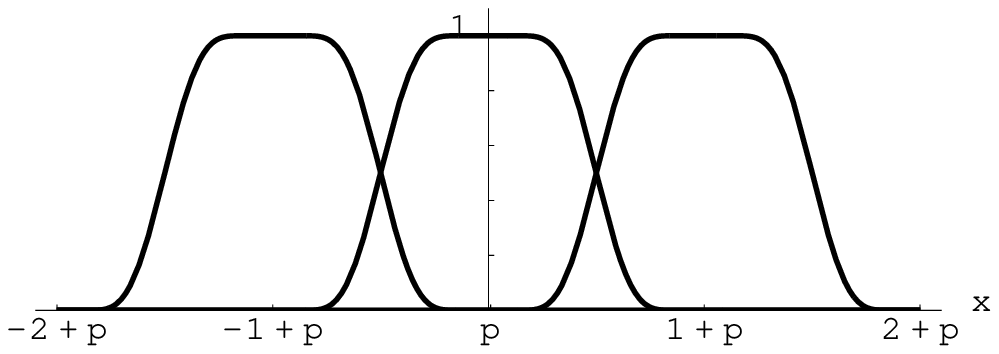}\\
 Fig.A1: A function $\varphi(x)$ decomposing unity.
\end{center}

The fourrier transform $\phi(k)$ of $\varphi(x)$ has the property
\begin{equation} \phi(0) = 1 \ \ , \ \ \phi(2p \pi) = 0  \ \ , \ \ \forall  p \ \ 
\mbox{integer} \ \ \neq 0 .\end{equation}
The coefficient $C_n$ in the expansion of $f(x)$ are then given by
\begin{equation} C_n = {1 \over \lambda} \int^{\infty}_{-\infty} f(x) \varphi({x \over
\lambda}) e^{inKx} dx. \end{equation}
If $f(x)$ is a standard integrable function of period $\lambda$, $C_n$ is just the usual
fourrier coefficient since $\varphi({x \over \lambda}) + \varphi({x \over \lambda} - 1)
 = 1$.

We consider now, for $(a,b) \in \mathbb R $, the distribution 
\begin{equation} T_{ab}(x) = {1 \over 2i} \sum^{\infty}_{p=-\infty}
{e^{i(ax + {b\over x})}\over x} 
 (1 - {\sin (\pi x) \over \pi x}) \delta(x-p). \end{equation}
With the $C^{\infty}$-test function $\Omega(x)$ which decomposes unity we have
%XXXXXXXXXXXXXXXXXXXXXXXXXXXX \\
\begin{eqnarray}
T_{ab}(x) (\Omega)  &=&  {1 \over 2i} \sum^{\infty}_{p=-\infty} \int^{\infty}_{-\infty}
 {e^{i(ax + {b\over x})}\over x} (1 - {\sin (\pi x) \over \pi x}) \delta(x-p)
\Omega(x) dx \nonumber \\
 &=&  {1 \over 2i} \sum^{\infty}_{p=-\infty} {e^{i(ap + {b\over p})}\over p}
(1 - {\sin \pi p \over \pi p}) \Omega(p) \nonumber\\
&=&   \sum^{\infty}_{p=1} {1 \over p}\sin (ap + {b \over p}) \end{eqnarray}
%XXXXXXXXXXXXXXXXXXX \\
since $\Omega(p) = 1 \ \ , \ \  \forall p$ integer or zero (cf Fig.A1).

On the other hand the periodic distribution
$$f(x) = \sum^{\infty}_{p = -\infty} \delta(x-p)$$
admits also the fourrier expansion (A.1) with $K = 2\pi$ and $C_n = 1$, directly from
(A.5). Hence we have the well known representation
\begin{equation}
\sum^{\infty}_{p=-\infty} e^{2ip\pi x} = \sum^{\infty}_{p=-\infty} \delta(x-p). 
\end{equation}
$T_{ab}(x)(\Omega)$ is then also given by

\begin{equation}
T_{ab}(x)(\Omega) = \sum^{\infty}_{p=-\infty} \int^{\infty}_{0} {dx \over x} \Omega(x)
\sin [(a+2p \pi)x + {b \over x}] (1 - {\sin \pi x \over \pi x}). \end{equation}
$\Omega(x)$ being a decomposition of unity and since the integral is well defined with 
$\Omega(x) = 1$ on the whole integration domain, we have
\begin{equation}
\int^{\infty}_0 {dx \over x} \sin  [(a+2p \pi)x + {b \over x}] = {\pi \over 2}
 [sign	(a+2p\pi) + sign(b)] J_0 (2 \sqrt{(a+2p \pi)b}), \end{equation}
and
$$ \int^{\infty}_0 {dx \over x} \sin [(a+2p \pi)x + {b \over x}] {\sin \pi x \over \pi x} =
{1\over 4b} \{[sign(a +(2p+1)\pi) + sign(b)]$$
$$\sqrt{(a+(2p+1)\pi)b} \ \ J_1(2 \sqrt{(a+
(2p+1)\pi)b}) 
- [sign(a+(2p-1)\pi)+sign(b)]$$
\begin{equation}\sqrt{(a+(2p-1)\pi)b} \ \ J_1(2 \sqrt{(a+(2p-1)\pi)b})\}.  \end{equation}
%%%%%%%%%%%%%%%%%%%%%%%%%
Here  $sign(x) = \pm 1 \ \ \ x  \gtrless 0 \ \ , \ \  sign(0) = 0$ and $J_n(x)$ is
the ordinary Bessel function of order $n$ .

Specializing to the discretized light-cone variables $a = {\pi x^- \over L}, b = {m^2
\over 4} . {L x^+ \over \pi}$, we have
$${2 \over \pi} \sum^{\infty}_{n=1} {1 \over n} \sin [{n \pi x^- \over L} + {m^2 \over
4} {Lx^+ \over n \pi}] = \sum^{\infty}_{p = - \infty} [sign (x^- + 2pL) + sign(x^+)] \
\ J_0
(m \sqrt{x^2+2pLx^+}) $$
$$- {1 \over mLx^+} \sum^{\infty}_{p = - \infty} \{[sign (x^- + (2p+1)L) + sign(x^+)]
\sqrt{x^2+(2p+1)Lx^+} \ \ J_1(m \sqrt{x^2 + (2p+1)Lx^+})$$
\begin{equation} - [sign(x^-+(2p-1)L) + sign(x^+)] \sqrt{x^2+(2p-1)L x^+} \ \ 
J_1(m \sqrt{x^2 + (2p-1)Lx^+})\}. \end{equation}

This is invariant indeed under the replacement $x^- \to x^- + 2mL \ \ , \ \ \forall m$ integer.
If $x^+ = 0$ one has, since $-L \leq x^- \leq L$, and $\forall N \ \  \mbox{integer} > 0$

\begin{equation} \sum^N_{p=-N} sign [x^-+2pL] = sign (x^-) \end{equation}
and
\begin{equation} {1 \over 2L} \sum^N_{p=-N} \{sign [x^- + (2p+1)L][x^-+(2p+1)L] - sign
[x^-+(2p-1)L][x^-+(2p-1)L]\} = {x^- \over L}, \end{equation}
in agreement with eq. (2.3).

For non zero $x^+$, (A.12) reduces to the continuum causal contribution of eq. (2.8) and
non-causal terms :
$$ \frac{2}{\pi} \sum^{\infty}_{n=1} \frac{1}{n} \sin [\frac{n\pi x^-}{L} +
\frac{m^2}{4} \frac{Lx^+}{n \pi}] = [sign(x^+) + sign(x^-)] J_0(m \sqrt{x^+x^-}) +$$
$$\lim_{N \to \infty} \left\{ 2 sign(x^+) \sum^N_{p=1} J_0(m \sqrt{x^+x^- + 2pLx^+ sign(x^+)})
\right. $$
\begin{equation}  
\left. - \frac{2}{mLx^+} \sqrt{x^+x^- +(2N+1) Lx^+ sign(x^+)} J_1(m 
\sqrt{x^+x^- +(2N+1) Lx^+ sign(x^+)}\right\}. \end{equation}

The limit $N \to \infty$ in (A.15) is still elusive because the compensation between the
two diverging terms in $N$ is not explicit. However the remaining sum in (A.15) can be
given in an integral form using the contour integral representation of $J_0(z)$

\begin{equation} J_0(z) = \frac{1}{2\pi i} \int_C {ds \over s} e^{(s - {z^2 \over 4s})} , 
\end{equation}
where $C$ is the contour aroud the negative real axis and encircling the origin in the clock-wise direction. Then the geometric sum over p can be performed and, with $\alpha = \frac{m^2}{4} x^+x^-$ and $\beta = \frac{m^2}{2} Lx^+ sign (x^+)$, positive, we have the result 

\begin{eqnarray}
&&\sum^N_{p=1} J_0(m \sqrt{x^+x^-+2pLx^+ sign(x^+)})  =
\frac{1}{2 \pi i} \int_C \frac{dz}{z} 
e^{(z - \frac{\alpha + \beta}{z})} 
[ \frac{1 - e^{-\frac{N\beta}{z}}}{1 - e^{- \frac{\beta}{z}}}] \nonumber \\
&& = \frac{1}{2 \pi i} 
\int_C \frac{dz}{z} e^{\sqrt{\alpha + \frac{\beta}{2}(N+1)} (z-\frac{1}{z})}
\frac{sinh(\frac{N \beta z}{2 \sqrt{\alpha + \frac{\beta}{2}(N+1)}})}
{sinh(\frac{ \beta z}{2 \sqrt{\alpha + \frac{\beta}{2}(N+1)}})} \nonumber \\ 
&& = \frac{1}{\beta} (\sqrt{\alpha + \beta (N+\frac{1}{2})} J_1(2 \sqrt{\alpha + \beta (N+\frac{1}{2})})-\sqrt{\alpha +\frac{\beta}{2}} J_1(2 \sqrt{\alpha +\frac{\beta}{2}}))  
\end{eqnarray}
since the hyperbolic sine in the denominator reduces to its argument in the large N limit. Collecting terms in (A.17) we have the result
%\vfill\eject

$$2 sign (x^+) \sum^{N}_{p=1} J_0 (m \sqrt{x^+x^- + 2pLx^+ sign(x^+)}) = - {2 \over
mLx^+} \sqrt{x^+x^- + Lx^+ sign(x^+)}$$
$$J_1 (m \sqrt{x^+x^- + Lx^+ sign(x^+)})  +{2 \over mLx^+} 
\sqrt{x^+x^- + (2N+1)Lx^+ sign(x^+)} $$
\begin{equation}  
J_1 (m \sqrt{x^+x^- + (2N+1)Lx^+ sign(x^+)}) + O(L^{-5/4}) \end{equation}

Now the limit $N \to \infty$ can be taken in (A.15) as the diverging term in $N$ in 
(A.15) is cancelled exactly by the one in (A.18), leaving the result

$${2 \over \pi} \sum^{\infty}_{n = 1} {1 \over n} \sin ({n \pi x^- \over L} + {m^2
\over 4} {Lx^+ \over n \pi}) = [sign (x^+) + sign (x^-)] J_0(m \sqrt{x^+x^-)}$$
\begin{equation}
- {2 \over mLx^+} \sqrt{x^+x^- + 2Lx^+ sign(x^+)} 
J_1 (m \sqrt{x^+x^- + 2Lx^+ sign(x^+)}) + O(L^{-5/4}).\end{equation}

\end{document}